\documentclass[longauth,letter,traditabstract]{aa}

\usepackage{graphicx}
\usepackage{txfonts}
\begin{document}
   \title{HerMES: Far-infrared properties of known AGN in the HerMES fields\thanks{{\it Herschel} is an ESA space observatory with science instruments provided by European-led Principal Investigator consortia and with important participation from NASA}}

\author{E.~Hatziminaoglou\inst{1}
\and A.~Omont\inst{2}
\and J.\,A.~Stevens\inst{3}
\and A.~Amblard\inst{4}
\and V.~Arumugam\inst{5}
\and R.~Auld\inst{6}
\and H.~Aussel\inst{7}
\and T.~Babbedge\inst{8}
\and A.~Blain\inst{9}
\and J.~Bock\inst{9,10}
\and A.~Boselli\inst{11}
\and V.~Buat\inst{11}
\and D.~Burgarella\inst{11}
\and N.~Castro-Rodr{\'\i}guez\inst{12}
\and A.~Cava\inst{12}
\and P.~Chanial\inst{8}
\and D.\,L.~Clements\inst{8}
\and A.~Conley\inst{13}
\and L.~Conversi\inst{14}
\and A.~Cooray\inst{4,9}
\and C.\,D.~Dowell\inst{9,10}
\and E.~Dwek\inst{15}
\and S.~Dye\inst{6}
\and S.~Eales\inst{6}
\and D.~Elbaz\inst{7}
\and D.~Farrah\inst{16}
\and M.~Fox\inst{8}
\and A.~Franceschini\inst{17}
\and W.~Gear\inst{6}
\and J.~Glenn\inst{13}
\and E.\,A.~Gonz\'alez~Solares\inst{18}
\and M.~Griffin\inst{6}
\and M.~Halpern\inst{19}
\and E.~Ibar\inst{20}
\and K.~Isaak\inst{6}
\and R.\,J.~Ivison\inst{20,5}
\and G.~Lagache\inst{21}
\and L.~Levenson\inst{9,10}
\and N.~Lu\inst{9,22}
\and S.~Madden\inst{7}
\and B.~Maffei\inst{23}
\and G.~Mainetti\inst{17}
\and L.~Marchetti\inst{17}
\and A.\,M.\,J.~Mortier\inst{8}
\and H.\,T.~Nguyen\inst{9,10}
\and B.~O'Halloran\inst{8}
\and S.\,J.~Oliver\inst{16}
\and M.\,J.~Page\inst{24}
\and P.~Panuzzo\inst{7}
\and A.~Papageorgiou\inst{6}
\and C.\,P.~Pearson\inst{25,26}
\and I.~P{\'e}rez-Fournon\inst{12}
\and M.~Pohlen\inst{6}
\and J.\,I.~Rawlings\inst{24}
\and D.~Rigopoulou\inst{25,27}
\and D.~Rizzo\inst{8}
\and I.\,G.~Roseboom\inst{16}
\and M.~Rowan-Robinson\inst{8}
\and M.~Sanchez Portal\inst{14}
\and B.~Schulz\inst{9,22}
\and Douglas~Scott\inst{19}
\and N.~Seymour\inst{24}
\and D.\,L.~Shupe\inst{9,22}
\and A.\,J.~Smith\inst{16}
\and M.~Symeonidis\inst{24}
\and M.~Trichas\inst{8}
\and K.\,E.~Tugwell\inst{24}
\and M.~Vaccari\inst{17}
\and I.~Valtchanov\inst{14}
\and L.~Vigroux\inst{2}
\and L.~Wang\inst{16}
\and R.~Ward\inst{16}
\and G.~Wright\inst{20}
\and C.\,K.~Xu\inst{9,22}
\and M.~Zemcov\inst{9,10}}

\institute{ESO, Karl-Schwarzschild-Str.\ 2, 85748 Garching bei M\"unchen, Germany\\
 \email{ehatzimi@eso.org}
\and Institut d'Astrophysique de Paris, UMR 7095, CNRS, UPMC Univ.\ Paris 06, 98bis boulevard Arago, F-75014 Paris, France
\and Centre for Astrophysics Research, University of Hertfordshire, College Lane, Hatfield, Hertfordshire AL10 9AB, UK
\and Dept.\ of Physics \& Astronomy, University of California, Irvine, CA 92697, USA
\and Institute for Astronomy, University of Edinburgh, Royal Observatory, Blackford Hill, Edinburgh EH9 3HJ, UK
\and Cardiff School of Physics and Astronomy, Cardiff University, Queens Buildings, The Parade, Cardiff CF24 3AA, UK
\and Laboratoire AIM-Paris-Saclay, CEA/DSM/Irfu - CNRS - Universit\'e Paris Diderot, CE-Saclay, pt courrier 131, F-91191 Gif-sur-Yvette, France
\and Astrophysics Group, Imperial College London, Blackett Laboratory, Prince Consort Road, London SW7 2AZ, UK
\and California Institute of Technology, 1200 E.\ California Blvd., Pasadena, CA 91125, USA
\and Jet Propulsion Laboratory, 4800 Oak Grove Drive, Pasadena, CA 91109, USA
\and Laboratoire d'Astrophysique de Marseille, OAMP, Universit\'e Aix-marseille, CNRS, 38 rue Fr\'ed\'eric Joliot-Curie, 13388 Marseille cedex 13, France
\and Instituto de Astrof{\'\i}sica de Canarias (IAC) and Departamento de Astrof{\'\i}sica, Universidad de La Laguna (ULL), La Laguna, Tenerife, Spain
\and Dept.\ of Astrophysical and Planetary Sciences, CASA 389-UCB, University of Colorado, Boulder, CO 80309, USA
\and Herschel Science Centre, European Space Astronomy Centre, Villanueva de la Ca\~nada, 28691 Madrid, Spain
\and Observational  Cosmology Lab, Code 665, NASA Goddard Space Flight  Center, Greenbelt, MD 20771, USA
\and Astronomy Centre, Dept.\ of Physics \& Astronomy, University of Sussex, Brighton BN1 9QH, UK
\and Dipartimento di Astronomia, Universit\`{a} di Padova, vicolo Osservatorio, 3, 35122 Padova, Italy
\and Institute of Astronomy, University of Cambridge, Madingley Road, Cambridge CB3 0HA, UK
\and Department of Physics \& Astronomy, University of British Columbia, 6224 Agricultural Road, Vancouver, BC V6T~1Z1, Canada
\and UK Astronomy Technology Centre, Royal Observatory, Blackford Hill, Edinburgh EH9 3HJ, UK
\and Institut d'Astrophysique Spatiale (IAS), b\^atiment 121, Universit\'e Paris-Sud 11 and CNRS (UMR 8617), 91405 Orsay, France
\and Infrared Processing and Analysis Center, MS 100-22, California Institute of Technology, JPL, Pasadena, CA 91125, USA
\and School of Physics and Astronomy, The University of Manchester, Alan Turing Building, Oxford Road, Manchester M13 9PL, UK
\and Mullard Space Science Laboratory, University College London, Holmbury St.\ Mary, Dorking, Surrey RH5 6NT, UK
\and Space Science \& Technology Department, Rutherford Appleton Laboratory, Chilton, Didcot, Oxfordshire OX11 0QX, UK
\and Institute for Space Imaging Science, University of Lethbridge, Lethbridge, Alberta, T1K 3M4, Canada
\and Astrophysics, Oxford University, Keble Road, Oxford OX1 3RH, UK}

   \date{}

\abstract{Nuclear and starburst activity are known to often occur concomitantly.
{\it Herschel}-SPIRE provides sampling of
the far-infrared (FIR) spectral energy distributions (SEDs) of type 1 and type 2 AGN, allowing for the
separation between the hot dust (torus) and cold dust (starburst) emission.
We study large samples of spectroscopically confirmed type
1 and type 2 AGN lying within the {\it Herschel} Multi-tiered Extragalactic Survey
(HerMES) fields observed during the science demonstration phase,
aiming to understand their FIR colour distributions
and constrain their starburst contributions.
We find that one third of the spectroscopically confirmed AGN in the HerMES
fields have 5$\sigma$ detections at 250\,$\mu$m, in agreement with previous (sub)mm AGN studies.
Their combined {\it Spitzer}-MIPS and {\it Herschel}-SPIRE colours (specifically $S_{250}/S_{70}$ vs
$S_{70}/S_{24}$) quite clearly separate them from the non-AGN, star
forming galaxy population,
as their 24 $\mu$m flux is dominated by the hot torus emission.
However, their SPIRE colours alone do not differ from
those of non-AGN galaxies. SED fitting shows that all those AGN
need a starburst component to fully account for their FIR emission.
For objects at $z>2$ we find a correlation between the infrared luminosity attributed to the starburst
component, $L_\mathrm{SB}$, and the AGN accretion luminosity, $L_\mathrm{acc}$, with
$L_\mathrm{SB} \propto L_\mathrm{acc}^{0.35}$.
Type 2 AGN detected at 250\,$\mu$m show on average higher $L_\mathrm{SB}$ than type 1 objects 
but their number is still too low to establish whether this trend indicates stronger star
formation activity.}


   \maketitle
%

\section{Introduction}

Active galactic nuclei (AGN) and starburst activity, both among the most energetic 
extragalactic phenomena, have been studied separately for decades.
However, it is only in the past decade and a half, with the advent 
of the infrared (IR) observatories (ISO, {\it Spitzer} and now 
{\it Herschel}; \cite{pilbratt10}) that it has become clear that the two phenomena
are related and, more often than not, happen concomitantly (e.g.
\cite{schweitzer06}).

The {\it Spitzer} IRAC and MIPS cameras sampled the spectral energy
distribution (SED) of extragalactic sources exactly where the peak of the
AGN dust emission is expected to occur under the paradigm of an axisymmetric
dust distribution (often referred to as {\it torus})
surrounding the central super-massive black hole. The IRS
spectrograph provided the necessary details of the silicate
feature in emission and in absorption and 
we now have a better understanding of the physics 
of hot dust around AGN (e.g.\ \cite{hao07}; \cite{levenson07}).
The peak of the cold dust emission, however, a tracer of star formation, was beyond the 
wavelength range explored by {\it Spitzer} for high redshift sources
and it is only now, with the advent of SPIRE (\cite{griffin10}), that
the full FIR SEDs of galaxies can be built, all the way to 500\,$\mu$m.

The work presented here intends to build on the previous experience
gathered with {\it Spitzer} and BLAST. 
We study large samples of spectroscopically
confirmed type 1 and type 2 AGN in the largest {\it Herschel} Multi-tiered Extragalactic
Survey (HerMES\footnote{http://hermes.sussex.ac.uk}; Oliver et al.\ in prep.) 
fields observed during the science demonstration phase (SDP).
The SEDs and IR properties of many of the objects in the samples have been
studied in the past (e.g.\ \cite{richards06}; Hatziminaoglou et al.\ 2008, 2009) 
and their AGN properties are well constrained.
The idea is to extend such study to the larger wavelengths now observed
by SPIRE in an effort to also constrain the starburst component of these
objects.


\section{AGN samples, their SPIRE detections and mid-to-FIR colours}
\label{sec:samples}

The AGN master sample considered here consists of a total of 469 spectroscopically
confirmed type 1 and type 2 AGN in the Lockman-SWIRE (LS) and the {\it Spitzer} First
Look Survey (FLS) fields, with redshifts that extend to $z>4$. 
More specifically, it includes SDSS quasars in LS and FLS; mid-infrared (MIR)
selected AGN in FLS with spectroscopic redshifts from the MMT-Hectospec 
(Papovich et al.\ 2006); two MIR-selected type 2 AGN samples from
Mart\'{\i}nez-Sansigre et al.\ (2006) and Lacy et al (2007); and two X-ray selected
type 2 samples from Mainieri et al.\ (2002) and Polletta et al.\ (2006).
Their SPIRE fluxes are estimated via linear inversion methods, using, whenever
available, the positions of known 24\,$\mu$m sources as priors (see Roseboom et al.\ in prep.).
The details of the various subsamples and the numbers of
250\,$\mu$m and additional 350\,$\mu$m 5$\sigma$ detections are given in 
Table \ref{tab:sample}. Note that none of the X-ray selected type 2 AGN from
Mainieri et al.\ (2002) were detected by SPIRE. 
All 5$\sigma$ detections at 250\,$\mu$m 
and 350\,$\mu$m have fluxes above 12.8 mJy and 12.2 mJy, respectively.
No flux cut is applied at 500\,$\mu$m, these fluxes are used, instead, at face value.
The different detection rates of the various samples in the two fields are the
result of the different depths (FLS being deeper than LS, Oliver et al.\ 2010b, this volume)
as well as the selections of the various samples.

\begin{table}[ht]
\caption{AGN sample reference, field,
type, number of objects, objects with 5$\sigma$ detections at 250 $\mu$m
and at both 250 and 350 $\mu$m.}
\label{tab:sample}
\centering          
\begin{tabular}{l c r  c}
\hline \hline
Sample & Type & Nobj & 5$\sigma$ 250\,$\mu$m\\
       &      &      & (and 350\,$\mu$m) \\
\hline
SDSS (LS) & 1 & 168 &  44 (26) \\
SDSS (FLS) & 1 & 86  &  29 (21) \\
\cite{papovich06} (FLS) & 1 & 159 &  71 (42) \\
\cite{lacy07} (FLS) & 2 & 20  &   5 (2) \\
\cite{martinez06} (FLS) & 2 & 16 & 5 (4)\\
\cite{polletta06} (LS) & 2 & 11  &   2 (1) \\
\cite{mainieri02} (LS) & 2 & 9   &   0 (0) \\
\hline                  
Total         &   & 469 & 156 (96) \\
\hline\hline 
\end{tabular}
\end{table}

It is worth noting that one third of the objects have 5$\sigma$ detections
at 250\,$\mu$m, also reported by \cite{elbaz10}.
This detection rate is in excellent agreement with results obtained from the studies
of bright ($M_B<-26.1$), high redshift ($z \geq 1.8$) quasars at 1.2 mm, with the MAx-Planck Millimeter
BOlometer (MAMBO) array at IRAM 30-metre telescope (\cite{carilli01}; \cite{omont01}).
It is, however, higher than the $\sim$15\%, reported by \cite{priddey03} for $1.5 < z < 3.0$ 
radio-quiet, luminous ($M_B<-27.5$)
quasars, observed at 850\,$\mu$m with SCUBA on the James Clerk Maxwell Telescope (JCMT).
Even though our derived detection rate holds for the whole absolute magnitude range of our sample
($-19 > M_g > -28.3$) as well as in broad $M_g$ bins (here we use $Mg$, as the SDSS $g$-filter is the
closest to the B-band used in the other studies), if we look at the more rare, brightest objects ($M_g<-27.5$)
we only find about 20\% of them having 5$\sigma$ detections at 250\,$\mu$m, close to the 15\% reported by
\cite{priddey03}.

Though it is difficult to say anything about the significance of the detections
of the rest of the objects 
in an objective way, the observed properties of the AGN with and without SPIRE
counterparts
are quite similar, both in terms of optical and IR fluxes and in redshift.
Therefore, since the present study only makes use of objects with 5$\sigma$ detections
at 250\,$\mu$m, and since 250\,$\mu$m emission arises from cold dust in the star forming regions,
the 250\,$\mu$m-detected sample may be biased towards objects with stronger star formation activity.
The issue of SPIRE detection rates of AGN is addressed in more detail in Stevens et al.\ in prep.

\begin{figure}
\centerline{
\includegraphics[width=8cm]{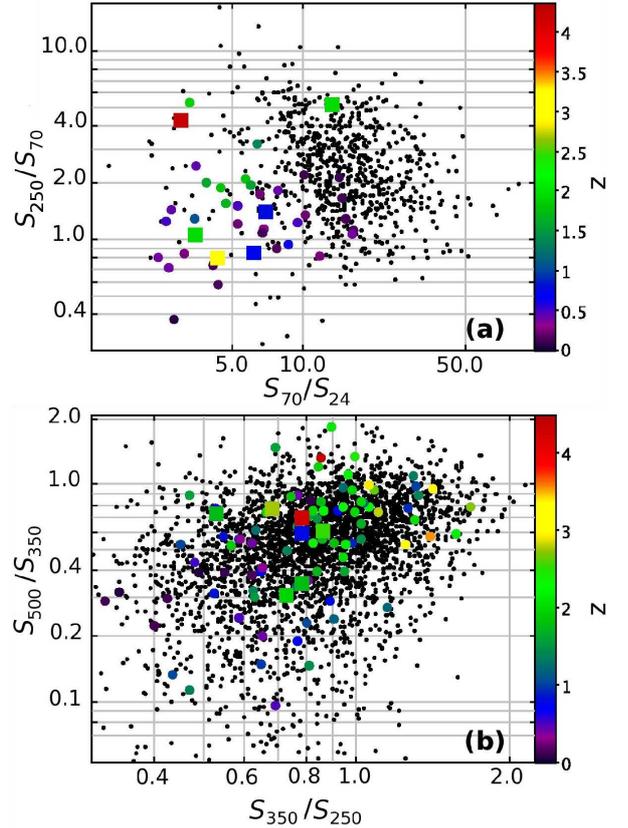}}
\caption{MIPS-SPIRE colours (a) and SPIRE colours (b)
for type 1 (circles) and type 2 (large squares) AGN as a function of redshift, 
over-plotted on the entire SPIRE catalogue in the FLS field,
for objects with 5$\sigma$ detections at 250$\,\mu$m (a) and 350\,$\mu$m (b).}
\label{fig:spirecol}
\end{figure}

Figure \ref{fig:spirecol} shows the MIPS-SPIRE colours $S_{250}/S_{70}$ vs 
$S_{70}/S_{24}$ (a) and SPIRE colours $S_{500}/S_{350}$
vs $S_{350}/S_{250}$ (b) of all AGN with 250\,$\mu$m and 350\,$\mu$m detections
above 5$\sigma$ (type 1: circles; type 2: squares) colour-coded
based on their redshift, and compared to all SPIRE sources in the
FLS field 
at the same SPIRE detection limits (black dots). AGN have much
bluer colours and tend to separate nicely from the
bulk of star forming non-active galaxies, as their 24\,$\mu$m fluxes are
dominated by the hot dust (torus) emission.
This is in agreement with recent MIR spectroscopic studies with {\it Spitzer}/IRS 
showing that the MIR quasar emission mostly arises from the torus,
independent on their FIR properties (\cite{netzer07}). The presence of strong
PAH features, however, may slightly weaken this effect in certain redshift ranges.
The 7.7\,$\mu$m PAH feature, in particular, might affect the 24\,$\mu$m flux of
$z \sim$ 2 AGN, when present (\cite{lutz08}).

In the SPIRE bands the colours of AGN are indistinguishable from 
those of the star forming non-active galaxies,
suggesting that AGN actually appear like starburst galaxies in the FIR.
This is consistent with longer wavelength studies that show the FIR 
emission of submm-luminous quasars and their starburst to share many properties 
of submm galaxies. Their FIR luminosities
are similar, and the best evidence is probably that they follow practically the 
same FIR/CO luminosity relation as submm galaxies and local ULIRGs (see e.g.\ 
Fig. 5 of \cite{riechers06}).

A variety of multi-wavelength datasets covering the energy range from the
radio to the X-rays is available for the objects in our master sample,
but in order to have a uniform wavelength SED coverage to perform the
fitting, we only used the available photometry from SDSS DR7 (\cite{abazajian09}),
2MASS and 2MASSx6 in the Lockman field (\cite{beichman03}), {\it Spitzer} IRAC and MIPS
data from the SWIRE (\cite{lonsdale04}) and FLS {\it Spitzer} Surveys, and 
new SPIRE HerMES data.
PACS (\cite{poglitsch10}) data have not been used, as they are only available for parts
of the HerMES SDP fields.


\section{Star formation in AGN}

The observed SEDs, described in Sect. \ref{sec:samples}, were compared,
by means of SED fitting with a standard $\chi^2$ minimisation, to
a series of models comprising three different components: a stellar component
composed by various simple stellar population (SSP) models build using the
Padova evolutionary tracks (\cite{bertelli94}); a grid of AGN/torus
models that include both a toroidal and a flared disk dust geometry
presented in \cite{fritz06}; and
six empirical starburst SEDs. For a full description of the 
SED fitting and individual model components see
Hatziminaoglou et al.\ (2008, 2009).
For reasons explained in detail in \cite{hatzimi08}
the reduced $\chi^2$ can reach high values without undermining our confidence
in the fits.  We will restrict the present study to the objects with 
fits having reduced $\chi^2<10$.

\begin{figure}
\centerline{
\includegraphics[width=7.5cm]{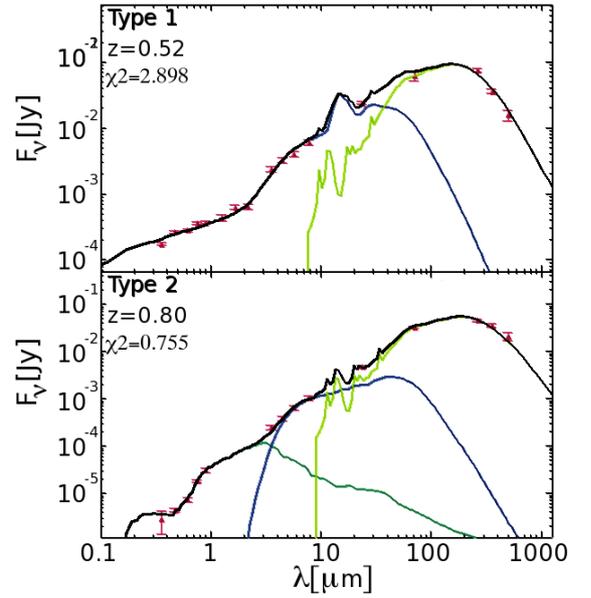}}
\caption{Example fits for a type 1 (upper panel) and a type 2 (lower panel) 
AGN. In red the observed
data points; in blue the AGN/torus model; in light green the starburst component;
in dark green the stellar component, and
in black the total model SED. The reported $\chi^2$ are {\it reduced}.}
\label{fig:examplefits}
\end{figure}

This leaves a total of 68 (42) type 1 and 11 (7) type 2
AGN, with 5$\sigma$ detections at 250\,$\mu$m (and 350\,$\mu$m), respectively.
As already mentioned, the 500\,$\mu$m are taken at face value, even though
the detection level of about one third of them 
falls below 2$\sigma$. The number drops to about 15\% for the objects with a 5$\sigma$ detection
at 350\,$\mu$m. Despite the low significance of some of the 500\,$\mu$m data points, they
follow nicely the observed SEDs as traced by the other FIR points and are unlikely to affect the
fit, because of their large photometric errors. SPIRE 500\,$\mu$m non-detections are not treated as upper
limits and are excluded from the fits.
Examples fits for a type 1 and a type 2 AGN are shown in Fig. \ref{fig:examplefits}.

As a first remark we point out that, in order to reproduce the observed
SPIRE data points, a starburst template is always needed, even if
we allow for very large (kpc-scale) tori. Large tori are, in any case,
unphysical in the sense that they extend well into the host galaxy where
other physical phenomena such as star formation may occur, and the
AGN is no longer the primary source of dust heating.
The SED fitting results in the set of estimations of physical 
parameters, describing the various components. Here we will focus on
the accretion luminosity, $L_\mathrm{acc}$, the model luminosity of the accretion disk ranging
from soft X-rays to the optical wavelengths scaled to the observed data points,
and the IR luminosity of the starburst component, $L_\mathrm{SB}$,
integrated between restframe 8 and 1000 $\mu$m, as a direct measure of star formation.

\begin{figure}
\centerline{
\includegraphics[width=8cm]{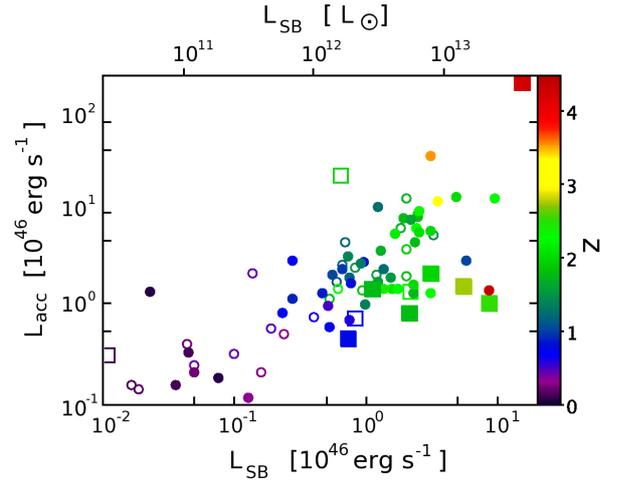}}
\caption{$L_\mathrm{SB}$ as a function of $L_\mathrm{acc}$ 
and redshift (colour-coding) for all
the objects (type 1: circles; type 2: large squares) 
with 5$\sigma$ detections at 250\,$\mu$m and 350\,$\mu$m
(filled symbols) and with 5$\sigma$ detections at 250\,$\mu$m alone
(additional open symbols).}
\label{fig:lsbdepend}
\end{figure}

Figure \ref{fig:lsbdepend}
shows $L_\mathrm{SB}$ as a function of $L_\mathrm{acc}$
with each point colour-coded according to redshift
for 250\,$\mu$m 5$\sigma$ detections alone (open symbols) and additional
350\,$\mu$m detections (filled symbols).
A broad correlation of $L_\mathrm{SB}$ can be seen with $L_\mathrm{acc}$, 
both quantities, however, also scale with redshift, as seen
from the colour-coding of the points. 

If we divide the sample in bins of redshift, the picture becomes less clear.
Table \ref{tab:corr} shows the Pearson correlation coefficients for the quantities 
$L_\mathrm{SB}$ and $L_\mathrm{acc}$ in five redshift bins.
(the missing value in the last column reflects the very small number of points
in the relevant bin, for which a Pearson correlation would be meaningless).
The very low values of the coefficients suggest that the observed global trend
may reflect the fact that the sample being flux limited, 
more distant objects will also be intrinsically more luminous.
A lack of correlation between $L_\mathrm{SB}$ and $L_\mathrm{acc}$ would
imply that star formation activity is not
influenced by the presence of an active nucleus in the centre of the galaxy.
Nevertheless, since the derived $L_\mathrm{acc}$ only covers at most two orders of 
magnitude at any given redshift, and since we are dealing with a limited number
of objects, it may be that our sample is not adequate to
detect any such correlations. 
Considering the $z>2.0$ bin alone we do find a 95\% (90\%) probability
of $L_\mathrm{SB}$ and $L_\mathrm{acc}$ correlating as $L_\mathrm{SB} \propto L_\mathrm{acc}^{0.35}$,
for objects with 5$\sigma$ detections at 250\,$\mu$m (and additionally at 350\,$\mu$m).
This value is in surprisingly good agreement
with that found by \cite{wang08} in their study of $z\sim$6 quasars with MAMBO.

\begin{table}[ht]
\caption{Correlation coefficients $r$ and $r'$ between $L_\mathrm{SB}$ and $L_\mathrm{acc}$
in bins of objects
considering the subsamples with 5$\sigma$ detections at 250 $\mu$m alone (N)
and those with both 250 and 350 $\mu$m 5$\sigma$ detections (N'), respectively.}
\label{tab:corr}
\centering
\begin{tabular}{r r r c c}
\hline \hline
$z$    & N &  N' &  $r$ & $r'$ \\
\hline
$z<0.5$     & 15 &  4  & 0.063 & -- \\
$0.5<z<1.0$ & 14 & 10  & 0.073 & 0.057 \\
$1.0<z<1.5$ &  9 &  6  & 0.054 & 0.241 \\
$1.5<z<2.0$ & 21 & 13  & 0.109 & 0.241 \\
$z>2.0$     & 18 & 16  & 0.496 & 0.468 \\
\hline\hline
\end{tabular}
\end{table}

\begin{figure}
\centerline{
\includegraphics[width=6.5cm]{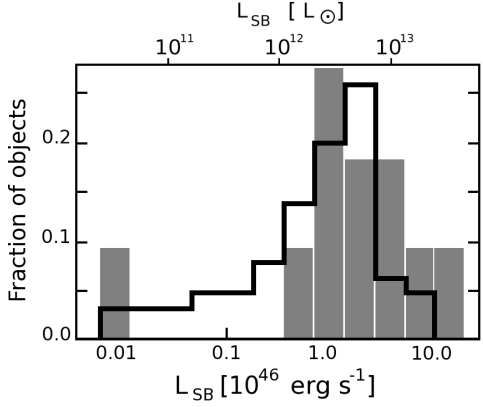}}
\caption{Distribution of
$L_{SB}$ (integrated between 8 and 1000 $\mu$m), as an indicator of star formation,
for type 1 (black) and type 2 (grey) AGN with 5$\sigma$
detections at 250\,$\mu$m.}
\label{fig:histolsb}
\end{figure}

An important issue in the study of nuclear and star formation activities is that of the
occurrence of star formation in the different types of AGN. 
Figure \ref{fig:histolsb} shows the distribution of $L_{SB}$, as an indicator of
star formation activity, for type 1 (black) and type 2 (grey) AGN for objects
with 5$\sigma$ detections at 250\,$\mu$m. Because of the difference in
number of AGN per type, the histogram shows the fraction of
objects per type.

Type 2 AGN detected at 250\,$\mu$m tend to have on average a larger $L_\mathrm{SB}$, trend also
observed in Fig. \ref{fig:lsbdepend} but their numbers
are too low to obtain a more robust estimate of this trend and to study
the possible implications.
The different criteria applied for the selection of type 1 and type 2 AGN 
(see Sect. \ref{sec:samples}), however, are rather unlikely to affect this result,
as none of the criteria particularly favours objects with stronger star formation:
the SDSS quasars and the X-ray selected type 2 AGN are
biased, if anything, towards AGN-dominated objects; the Lacy at al.\ (2007) type 2 AGN sample consists
of objects occupying the AGN-dominated part of the $S_{8.0}/S_{4.5}$ vs $S_{5.8}/S_{3.6}$ colour diagram
(Hatziminaoglou et al.\ 2008); finally, the MIR selected type 1 and type 2
AGN have been chosen among 24\,$\mu$m emitters, but as was already shown in Fig. \ref{fig:spirecol} 
the 24\,$\mu$m flux of AGN is dominated by the torus component.

\section{Conclusions}

Until recently, the study of type 1 and type 2 objects focused mainly
on their X-ray-to-MIR emission, with the two extreme $\lambda$ ranges sampling the
nuclear activity (X-rays) and hot dusty torus (MIR) emission, respectively, and the
middle wavelengths (optical-to-near IR) defining whether an AGN would
be classified as type 1 (unobscured) or type 2 (obscured).
With {\it Herschel}-SPIRE we can now also accurately probe 
the cold dust component, heated by star formation, and look for differences
between the types of active galaxies not only in the immediate environments of their
nuclei but also in their hosts.
The somewhat surprising lack of a strong correlation between the AGN intrinsic properties
($L_\mathrm{acc}$) and the star formation activity as traced by $L_\mathrm{SB}$,
at least up to $z\sim2$, possibly an effect of the narrow span of our sample in $L_\mathrm{acc}$, 
requires further investigation with larger AGN samples that also cover
a wider accretion luminosity range, including stacking sources undetected by SPIRE.
The same holds for the seemingly higher star formation activity in type 2 AGN. 
Even though this may not necessarily conflict with geometric unification schemes, it certainly opens
the way for investigations in new directions.

\begin{acknowledgements}
SPIRE has been developed by a consortium of institutes led by Cardiff University (UK) and including Univ.
Lethbridge (Canada); NAOC (China); CEA, LAM (France); IFSI, Univ. Padua (Italy); IAC (Spain);
Stockholm Observatory (Sweden); Imperial College London, RAL, UCL-MSSL, UKATC, Univ. Sussex
(UK); and Caltech, JPL, NHSC, Univ. Colorado (USA). This development has been supported by national
funding agencies: CSA (Canada); NAOC (China); CEA, CNES, CNRS (France); ASI (Italy); MCINN
(Spain); Stockholm Observatory (Sweden); STFC (UK); and NASA (USA).
The data presented in this paper will be released through the Herschel  
Database in Marseille HeDaM\footnote{http://hedam.oamp.fr/HerMES}.
This work makes use of data taken with the {\it Spitzer} Space Telescope,
the Sloan Digital Sky Survey (http://www.sdss.org) and 
the Two Micron All Sky Survey 
(http://www.ipac.caltech.edu/2mass/overview/access.html).
This work made use of Virtual Observatory tools and services for 
catalogue searches, cross-correlation and plotting
namely TOPCAT (http://www.star.bris.ac.uk/$\sim$mbt/topcat/), and VizieR
(http://vizier.u-strasbg.fr/cgi-bin/VizieR). EH would like to thank Kambiz Fathi
and Jacopo Fritz for the very useful discussions. We thank the anonymous referee for the
very insightful comments.

\end{acknowledgements}

\end{document}